\documentclass{jfm}
\usepackage{graphicx}
\usepackage{amsmath}
\usepackage{tabularx}
\usepackage{color}
\usepackage{comment}
\usepackage[colorlinks,citecolor = blue, linkcolor=red,hyperindex,CJKbookmarks]{hyperref}

\newcommand\Rich{\mbox{\textit{Ri}}}            
\newcommand\Frou{\mbox{\textit{Fr}}}            

\usepackage{siunitx}
\newcolumntype{Y}{>{\centering\arraybackslash}X}
\newcolumntype{P}[1]{>{\centering\arraybackslash}p{#1}}

\shorttitle{Bouncing dynamics of oil drops in stratified media}
\shortauthor{J.G. Meijer, Y. Li, C. Diddens and D.  Lohse}

\title{On the rising and sinking motion of bouncing oil drops in strongly stratified liquids}

\author{Jochem G. Meijer\aff{1}, 
  Yanshen Li\aff{2}\corresp{\email{liyanshen@ucas.ac.cn}},
  Christian Diddens\aff{1},\\
 \and Detlef Lohse\aff{1,}\aff{3}\corresp{\email{d.lohse@utwente.nl}}}

\affiliation{\aff{1}Physics of Fluids group, Max-Planck Center Twente for Complex Fluid Dynamics,  Department of Science and Technology, Mesa+ Institute and J. M. Burgers Center for Fluid Dynamics, University of Twente, P. O. Box, 217, 7500 AE Enschede, The Netherlands.
\aff{2}School of Engineering Science, University of Chinese Academy of Sciences, Beijing 101408, PR China
\aff{3}Max Planck Institute for Dynamics of Self-Organization, Am Fassberg 17, 37077 Göttingen, Germany.}

\begin{document}
\maketitle

\begin{abstract}
When an immiscible oil drop is immersed in a stably stratified ethanol-water mixture, the Marangoni flow on the surface of the drop can experience an oscillatory instability, so that the drop undergoes a transition from levitating to bouncing. The onset of the instability and its mechanisms have been studied previously \citep{li2021marangoni, li2022marangoni}, yet the bouncing motion of the drop itself, which is a completely different problem, has not yet been investigated. Here we study how the bouncing characteristics (jumping height, rising and sinking time) depend on the control parameters (drop radius, stratification strength, drop viscosity). We first record experimentally the bouncing trajectories of drops of different viscosities in different stratifications. Then a simplified dynamical analysis is performed to get the scaling relations of the jumping height and the rising \& sinking times. The rising \& sinking time scales are found to depend on the drag coefficient of the drop $C_D^S$ in the stratified liquid, which is determined empirically for the current parameter space \citep{zhang2019core}. For low viscosity (\SI{5}{cSt}) oil drops the results on the drag coefficient match the ones from the literature \citep{yick2009enhanced, candelier2014history}. 
For high viscosity (\SI{100}{cSt}) oil drops the parameter space had not been explored and the drag coefficients are not readily available. Numerical simulations are therefore performed to provide external verification for the drag coefficients, which well match with the experimental results.
\end{abstract}


\section{Introduction}\label{sec:introduction}
When a drop is placed in a stably stratified liquid with a concentration gradient, a Marangoni flow is generated on the surface of the drop. The interplay between this Marangoni flow and gravity will make the drop levitate or bounce continuously \citep{li2019bouncing, li2021marangoni}. The transition from levitating to bouncing is caused by an oscillatory Marangoni instability \citep{li2021marangoni}, which has two different mechanisms depending on the drop viscosity \citep{li2022marangoni}. However, once the drop starts to bounce,  the motion of the drop acts back on the velocity field and the flow field around the drop becomes completely different, namely unstationary rather than stationary. Thus, to calculate the motion of the drop itself becomes a completely new problem, as compared to calculating the onset of this instability. Though the bouncing cycle has been briefly and qualitatively described before \citep{li2019bouncing}, many questions regarding the properties of the bouncing trajectory remain unanswered.

The problem is actually that of a drop moving inside a density stratification with the presence of Marangoni flow. This problem is very complicated since the Marangoni advection is coupled with the drop motion and diffusion. 

Before further diving into this topic, it is beneficial to first briefly review the relatively simpler and more basic problem of a solid particle moving through a stratified medium, which is common in natural environments and of great interest in various scientific fields. One example is marine snow, which plays a central role in the marine carbon cycle and understanding its delayed vertical motion due to stratification is essential for bio-geochemical processes \citep{prairie2013delayed, prairie2015delayed}. Another example is the motion of aerosols in the stratified atmosphere which is of significant importance to the Earth's climate system, since they scatter and absorb a considerable amount of radiation \citep{jacobson1999fundamentals, huneeus2012estimating}.  A third example is the formation and sinking of ice crystals in the stably stratified layering of the saturated salt water in the Dead Sea in winter \citep{burns2012sediment, burns2015sediment, sutherland2015clay}, which is further complicated by the growth of the crystals during the sinking process.
The complicated physics involved when a spherical particle moves through a {(sharply)} stratified medium drew the attention of fluid dynamicists and the problem has since then been studied analytically \citep{zvirin1975settling, candelier2014history, mehaddi2018intertial}, numerically \citep{torres2000flow, doostmohammadi2014numerical, lee2019sedimentation,zhang2019core}, experimentally \citep{srdic1999gravitational, hanazaki2009jets,abaid2004internal,camassa2022critical} or by a combination of the above \citep{yick2009enhanced,camassa2010first}.

Coming back to the problem of a drop moving inside stratified liquids with the presence of Marangoni flow: apart from the motion of the particle, now a Marangoni flow is coupled as well. There has been some research on this topic. For example, \cite{blanchette2012drops} studied the motion of a denser drop inside a sharply stratified liquid, where the length scale of the stratification is smaller than the drop size. It was found that the settling drop could bounce up due to Marangoni flow. \cite{mandel2020retention} studied the rising of lighter drops in a two-layer density stratification. Here the length scale of the stratification is much larger than the drop size, but the potential Marangoni force is only analysed as a side effect, because in that case it is weak (at most comparable to the drop's buoyancy) and only assists the rising of the lighter drop, which will rise anyway due to buoyancy. Other research either ignored the density gradient of the surrounding medium \citep{young1959motion, chen1996marangoni, leven1976effect} or did not consider the Marangoni effect \citep{bayareh2013rising, shaik2020drag}. It is also worth mentioning that the effects of surfactants on drop motion have also been studied \citep{levich1962physicochemical, leven1976effect, chen1996marangoni, martin2017simulations}. For a more extensive summary regarding the motion in stratified liquids the reader is referred to the reviews by \citet{magnaudet2020particles} and \citet{more2022motion}.
For a review on further physicochemical hydrodynamical phenomena of drops we refer to \citet{lohse2020physicochemical}. 

For the problem considered here, the Marangoni force plays a major role in determining the speed and direction of the drop and the length scale of the stratification is much larger than the size of the drop. In order to understand how in this case the drop's trajectory changes with the physical properties, such as the drop viscosity, the drop size and the concentration gradient, experiments are performed on low (\SI{5}{cSt}) and high (\SI{100}{cSt}) viscosity silicone oil drops in various stratified ethanol-water mixtures. A simplified dynamical model is developed to help to understand the bouncing trajectories. The drag coefficient of the drop $C_D^S$ in the stratified fluid is found to be the key parameter to determine the rising and sinking time scales.
We take its dependence on the parameters from literature.
For the low viscosity drops, the drag coefficient has been suggested by \cite{yick2009enhanced} and \cite{candelier2014history}. For the high viscosity oil drops, there is no existing research, so numerical simulations are performed to provide independent verifications for the drag coefficient. The drag coefficients thus obtained are found to agree well with the experimental results.

The paper is organized as follows. In section \ref{sec:expermental&methods} the experimental procedure and the numerical method are described. In section \ref{sec:characteristics} the general characteristics of the bouncing trajectory are described. After that, in section \ref{sec:analysis}, a simplified dynamical analysis is provided to derive predictions for the minimum and maximum bouncing positions, as well as the dominant time scale for the rising and sinking motion of the drop, in which the drag coefficient $C_D^S$ is found to be the key parameter. In section \ref{sec:maxandminheight} we compare the theoretical predictions regarding the minimum and maximum bouncing positions to our experimental observations, finding good agreement. In section \ref{sec:dragexplain} we briefly summarize the main results discussed in the literature regarding the drag coefficient on spherical objects in stratified media, the results of the rising and sinking time scales of low and high viscosities are discussed in sections \ref{sec:results5cst} and \ref{sec:results100cst}, respectively. The paper ends with conclusions and an outlook (section \ref{sec:conclusion}). 

\section{Experimental procedure \& numerical methods}\label{sec:expermental&methods}

\subsection{Experimental procedure}

\begin{figure}
  \centerline{\includegraphics[width=0.95\textwidth]{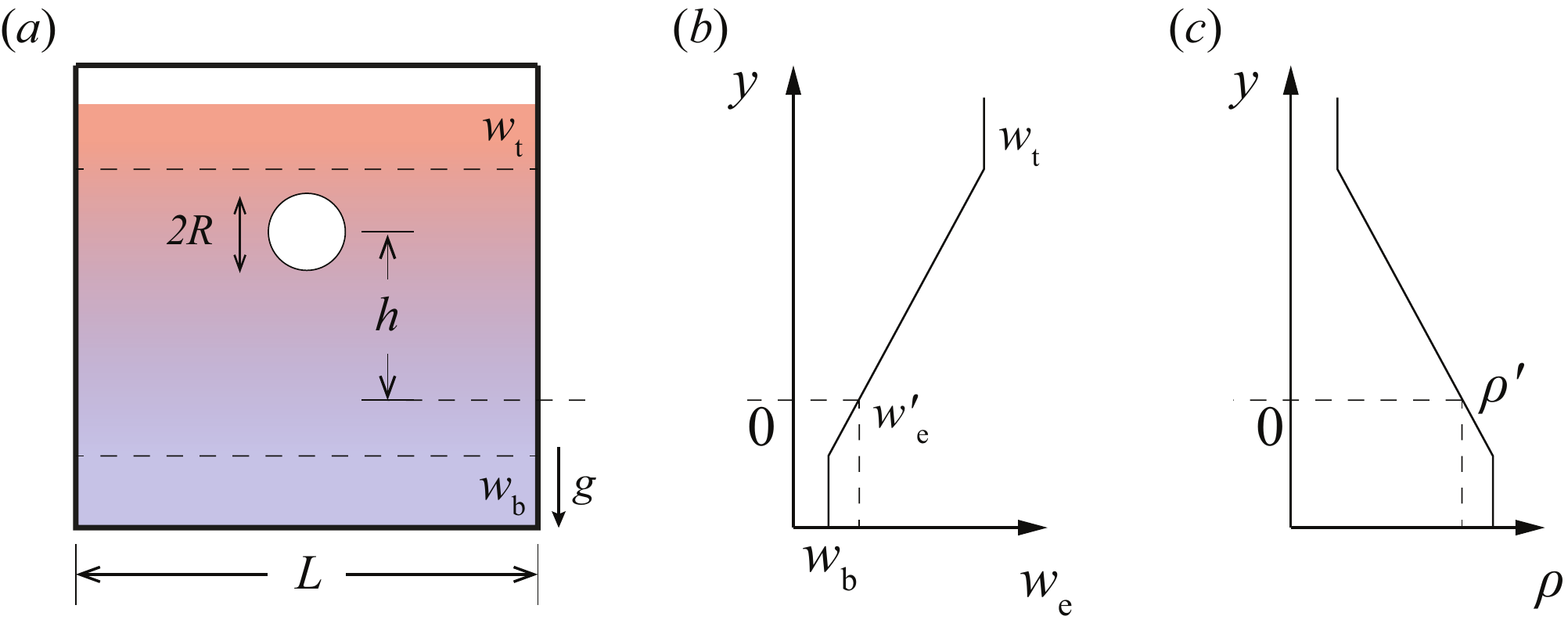}}
  \caption{($a$) Sketch of the experimental setup.  Using a modified version of the double-bucket method, a stable and linearly stratified ethanol-water mixture is generated in the middle of the container. Two liquid layers of uniform concentration are injected at the top ($w_{\mathrm{t}}$) and at the bottom ($w_{\mathrm{b}}$). The cubic glass container has an inner horizontal extension $L= \SI{30}{\milli\metre}$.  Silicone oil drops of varying radii $R$ and viscosities $\nu'$ are released from the top. ($b$) Ethanol weight fraction of the mixture $w_{\mathrm{e}}$ as a function of height.  ($c$) Density of the mixture $\rho$ as a function of height.  The density of the mixture matches that of the drop $\rho'$ at $y=0$.  This is called the density matched position where $\rho(w_{\mathrm{e}}')=\rho'$. The height of the drop $h$ is measured with respect to this density matched position.}
\label{fig:1}
\end{figure}

A sketch of the experimental setup is shown in figure \ref{fig:1}(\textit{a}).  A cubic glass container (Hellma, 704.001-OG, Germany) with inner horizontal extension $L= \SI{30}{\milli\metre}$ contains the linearly stratified ethanol-water mixture. The mixture is prepared using a ``double-syringe'' method \citep{li2022marangoni}, which is a slightly modified version of the double-bucket method \citep{oster1965density}. To avoid bubble formation during mixing, both ethanol (Boom B.V., 100 \%(v/v), technical grade, the Netherlands) and Milli-Q water are degassed in a desiccator at $\sim \SI{2000}{\pascal}$ for $\SI{20}{\min}$ before making the mixture.  Two layers of uniform ethanol concentration are located at the bottom (weight fraction $w_\mathrm{b}$) and at the top (weight fraction $w_{\mathrm{t}}$), between which the ethanol weight fraction $w_{\mathrm{e}}$ increases linearly, see figure \ref{fig:1}(\textit{b}).  Immediately after the mixture is prepared, $w_{\mathrm{e}}(y)$ is measured by laser deflection \citep{lin2013one,  li2019bouncing}. The two uniform layers $w_\mathrm{b}$ and $w_\mathrm{t}$ are used to  increase the accuracy of this method.  The density of the mixture $\rho(y)$ is calculated from $w_{\mathrm{e}}(y)$ using an empirical equation \citep{khattab2012density} and the height at which the density of the mixture $\rho(w_{\mathrm{e}}')$ matches the density of the oil $\rho'$ is set as $y = 0$, see figure \ref{fig:1}(\textit{c}). The values of $w_{\mathrm{b}}$ and $w_{\mathrm{t}}$, as well as the height of the stratified layer are varied to change the stratification strength $\mathrm{d}w_{\mathrm{e}}/\mathrm{d}y$.  \\
\indent
Drops are released from the top layer using a $\SI{1}{\micro\liter}$ syringe (Hamilton, KH7001) through an attached needle, whose outer diameter is $\SI{0.515}{\milli\metre}$.  They are released one at a time to ensure that only a single drop is present in the container. The properties of the different silicone oils (Sigma-Aldrich, Germany) are reported in table \ref{tab:properties}.  
Due to their robustness to surface contaminations \citep{young1959motion}, that alternatively would alter the interfacial surface properties and hence the motion of the drop, silicone oil droplets form the ideal candidate for our study.
The interfacial tensions $\sigma$ between both silicone oils and several ethanol-water mixtures are measured on a goniometer (OCA 15Pro, DataPhysics, Germany) by using the pendant-drop method, see Appendix \ref{appA}. 
The drop is illuminated by a collimated LED (Thorlabs, MWWHL4) and its motion is captured by a side-view camera (Nikon D850) connected to a long working distance lens (Thorlabs, MVL12X12Z plus 0.25X lens attachment).  All images are recorded at 30 frames per second.  After the drop has completed its third bouncing cycle, it is carefully taken out of the mixture using a second thin needle.  In all the cases, the third bouncing cycle is used to study its dynamics.  After this, another slightly smaller drop is released, whose motion is now captured. We repeat this process several times but for no longer than 40 minutes, as by that time diffusion will affect the linear stratification of the mixture. 
The stability of the background density gradient as a function of time has been discussed by \cite{li2019bouncing}, where it remained stable for more than an hour.
If still more data is desired, a new mixture is generated and the entire process is repeated.    \\
\indent
As the drop moves though the liquid, its motion will cause perturbations in the flow field that extend over a typical length scale.  Since this length scale depends on the stratification strength of the mixture \citep{phillips1970flows, wunsch1970oceanic},  it can happen that for weak density gradients it is comparable to or even larger than the actual size of the container.  The finite size of the container might therefore affect the motion of the drop as the stratifications become weaker \citep{li2021marangoni}. To exclude such container size effects, experiments with weaker density stratification i.e., $\mathrm{d}w_{\mathrm{e}}/\mathrm{d}y<\SI{60}{\per\metre}$ are performed in a larger container (Hellma, 704.003-OG, Germany) with an inner horizontal extension $L= \SI{50}{\milli\metre}$.  

\begin{table}
\vspace{-5mm} 
\begin{center}
\resizebox{\textwidth}{!}{\begin{tabular}{lccc}
\multicolumn{1}{c}{Liquid} & \multicolumn{1}{c}{\begin{tabular}[c]{@{}c@{}}Viscosity\\ $\nu'$ (cSt)\end{tabular}} & \multicolumn{1}{c}{\begin{tabular}[c]{@{}c@{}}Density\\ $\rho'$ ($\si{\kg\per\cubic\metre}$)\end{tabular}} & \multicolumn{1}{c}{\begin{tabular}[c]{@{}c@{}}Ethanol weight fraction $w_{\mathrm{e}}'$ at the density\\ matched position\end{tabular}} \\ \hline
5 cSt silicone oil         & 5                                                                                    & 913                                                                                      & 49.3 wt\%                                                                                                                    \\
100 cSt silicone oil       & 100                                                                                  & 966                                                                                      & 21.0 wt\%                                                                                                                    \\ \hline
\end{tabular}}
\vspace{-4mm} 
\caption{Properties of the silicone oils used in the experiments.}
\label{tab:properties}
\end{center}
\vspace{-1mm} 
\end{table}

\subsection{Numerical simulation}
\label{NumericalMethods}
Numerical simulations of the whole bouncing cycle are also performed to provide independent verification for the drag coefficients. 
The simulations utilize the numerical framework that has been developed to simulate the evaporation of multi-component drops \citep{diddens2017detailed}. 
Below, we briefly discuss its details. 

In order for the model to be compared to the experimental data it has to account for all the relevant physical mechanisms during the bouncing process. In particular, these are the flow driven by Marangoni and buoyancy effects and the advection and diffusion of the ethanol-water mixture outside the drop. Mass transfer across the drop's interface is not taken into account because the solubility of ethanol in silicone oil is negligible, or vice versa. The interface of the drop needs to be well-resolved to capture the Marangoni flow.  A sharp-interface finite element method has been developed, where the mesh is always conforming with the moving interface.  In addition, the mesh is treated as a pseudo-elastic body \citep{cairncross2000finite}, so that the bulk nodes follow the motion of the interfacial nodes. With moving mesh nodes, the numerical approach belongs to the class of arbitrary Eulerian-Lagrangian methods (ALE), which furthermore require us to consider the nodal movement $\dot{\mathbf{R}}$ at the interface. The problem is solved in axisymmetric cylindrical coordinates. 

The governing incompressible Navier-Stokes equations are:
\begin{equation}
\rho_0 \left( \partial _t \mathbf{u} + \mathbf{u} \bcdot \bnabla \mathbf{u} \right) = - \bnabla p^{\phi} - \rho^{\phi} \left(w_{\mathrm{e}} \right) g \mathbf{j} + \bnabla \bcdot \left[\mu^{\phi} \left(w_{\mathrm{e}} \right)  \left( \bnabla \mathbf{u} + \left(\bnabla \mathbf{u} \right)^{\mathrm{T}} \right) \right],
\label{eq:ns_momentum}
\end{equation}
\begin{equation}
\bnabla \bcdot \mathbf{u} = 0,
\end{equation}
where $\phi = \mathrm{d}, \mathrm{b}$ denotes the phase, i.e., the drop and the bulk liquid, respectively.
Using the Boussinesq approximation, the composition dependent mass density $\rho^{\mathrm{b}} \left(w_{\mathrm{e}} \right)$ is only considered for the gravitational term in Eq. (\ref{eq:ns_momentum}).  For the inertial term a constant density $\rho_0$ is assumed. 
The mass fraction $w_{\mathrm{w}}^{\mathrm{b}}$ for water is determined using the following advection-diffusion equation:
\begin{equation}
\partial _t w_{\mathrm{w}}^{\mathrm{b}} +  \mathbf{u} \bcdot \bnabla w_{\mathrm{w}}^{\mathrm{b}} = \bnabla\bcdot\left(D^\text{b}(w_\text{e})\bnabla w_\text{w}^\text{b}\right),
\end{equation}
with $D^{\mathrm{b}}  \left(w_{\mathrm{e}} \right)$ the composition dependent diffusivity in the bulk. 
The ethanol content is determined from the remainder of the water content.
To account for the high $\Pen$ numbers due to the low diffusivity, the equations are stabilised by a streamline upwind Petrov-Galerkin method (SUPG) \citep{brooks1982streamline}.
At the interface the boundary conditions are
\begin{equation}
\left( \mathbf{u} - \dot{\mathbf{R}} \right) \bcdot \mathbf{n} = 0,
\end{equation}
\begin{equation}
\boldsymbol{\tau}^{\mathrm{d}} \bcdot \mathbf{n} - \boldsymbol{\tau}^{\mathrm{b}} \bcdot \mathbf{n} = \sigma \left(w_{\mathrm{e}} \right) \kappa \mathbf{n} + \bnabla_{\mathrm{S}} \sigma \left(w_{\mathrm{e}} \right).
\end{equation}
Here, $\mathbf{n}$ is the unit interface normal pointing from the drop to the bulk liquid, $\boldsymbol{\tau}^{\phi}$ are the stress tensors in both phases, $\sigma \left(w_{\mathrm{e}} \right)$ is the composition dependent surface tension, $\kappa$ the curvature and $\bnabla_{\mathrm{S}}$ the surface gradient.
All composition dependent physical properties of the bulk liquid are implemented by a best fit of their corresponding values on the ethanol weight fraction, see Appendix \ref{appA}. The above equations are implemented in the finite element framework OOMP-LIB \citep{heil2006oomph} on triangular mixed first order Lagrange elements for the composition and conventional Taylor-Hood elements for the velocity and pressure.  To prevent the mesh from deforming significantly, it is reconstructed whenever the mesh quality falls below a specific threshold.  
Identical simulations are repeated with different mesh and domain sizes to ensure that the obtained results are not affected significantly. 

Results of the numerical simulations and their comparison to the experimental observations are discussed in more detail in section \ref{sec:results100cst}.  The sections that follow first focus on the experiments.

\section{General characteristics of the bouncing cycle}\label{sec:characteristics}
\begin{figure}
  \centerline{\includegraphics[width=\textwidth]{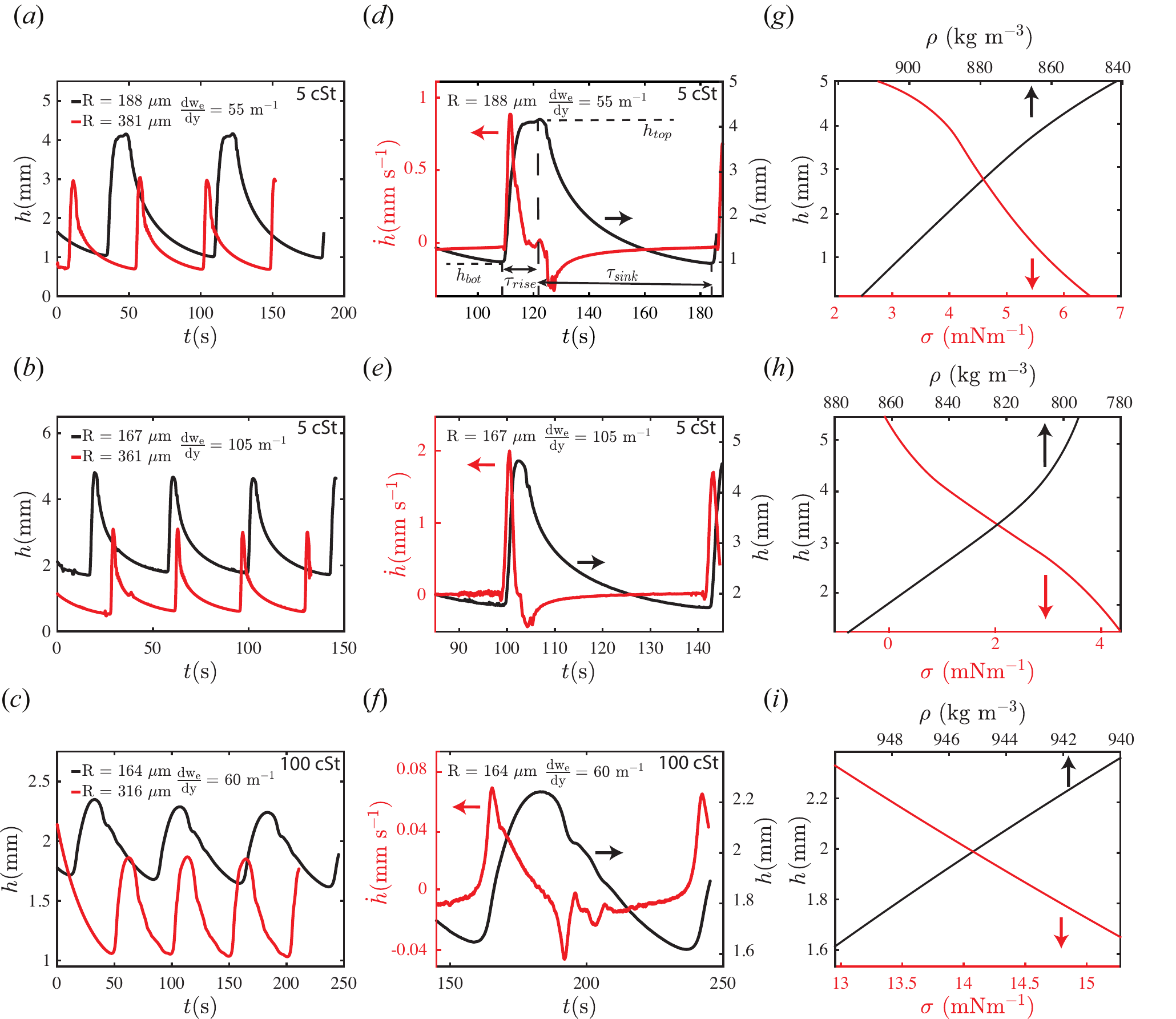}}
  \caption{($a$)-($b$) Experimentally obtained trajectories of \SI{5}{cSt} silicone oil drops of different radii in two different linearly stratified ethanol-water mixtures with $\mathrm{d}w_{\mathrm{e}}/\mathrm{d}y \approx \SI{55}{m^{-1}}$ and $\mathrm{d}w_{\mathrm{e}}/\mathrm{d}y \approx \SI{105}{m^{-1}}$, respectively.  At $t = \SI{0}{s}$ the tracking procedure is initialized. The density matched position, i.e., the position at which $\rho = \rho'$, is at $h = \SI{0}{mm}$. ($c$) Trajectories of two \SI{100}{cSt} silicone oil drops of different radii inside a linearly stratified ethanol-water mixture with $\mathrm{d}w_{\mathrm{e}}/\mathrm{d}y \approx \SI{60}{m^{-1}}$.  ($d$)-($f$) Velocity $\dot{h}$ and height $h$ as functions of time for a single bouncing cycle for the smaller drop inside the corresponding linearly stratified mixture. The definitions of some characteristics of the bouncing cycle are indicated in ($d$), namely the two extrema, $h_{\mathrm{top}}$ and $h_{\mathrm{bot}}$,  as well as the definition of the rising and sinking time intervals, $\tau_{\mathrm{rise}}$ and $\tau_{\mathrm{sink}}$. 
($g$)-($i$) Experimentally measured profiles of the mass density $\rho(y)$ and the interfacial surface tension $\sigma(y)$ using the laser deflection technique.}
\label{fig:2}
\end{figure}

When an immiscible drop is placed in a stably stratified ethanol-water mixture, the ethanol concentration gradient leads to a surface tension gradient on the surface of the drop, which points downwards, and a downward Marangoni flow is generated as a consequence. When the Marangoni flow is stable, the drop levitates at a fixed height; otherwise, for larger drops, the Marangoni flow is oscillatory due to an oscillatory instability \citep{li2022marangoni} and the drop bounces continuously. 
Whereas our previous studies looked into the onset of the bouncing instability, what mechanism triggers it and how it depends on the viscosity of the oil \citep{li2019bouncing,li2021marangoni,li2022marangoni}, here the dynamics of the bouncing cycle itself is studied.
Figure \ref{fig:2}($a$) shows the trajectories of two \SI{5}{cSt} drops of different radii in a stable stratification with $\mathrm{d}w_\mathrm{e}/\mathrm{d}y=\SI{55}{m^{-1}}$. As can be seen, the drops bounce periodically. Before the background gradient is changed due to diffusion and the bouncing of the drop, each period of the bouncing trajectory is identical. Figure \ref{fig:2}($d$) zooms in on one period of the trajectory of the $R=\SI{188}{\micro\meter}$ drop. The drop first sinks towards its density-matched position (before \SI{110}{\second}) due to gravity. During this period, an entrained liquid layer with almost uniform concentration is dragged down with the drop and the Marangoni flow on the drop is very weak \citep{li2019bouncing}. Also because of the enhanced drag caused by the stable stratification, the sinking velocity decreases exponentially \citep{zvirin1975settling, li2019bouncing}. Later, this entrained layer breaks due to diffusion and the Marangoni flow velocity increases exponentially \citep{li2019bouncing}, leading to the sudden upward jump of the drop (after $\sim\SI{110}{\second}$). 
Because of this exponential behaviour, the Marangoni flow velocity reaches its maximum value in a short time period. Later, as the drop moves up, the buoyancy force on it increases, so it decelerates, until a time when the drop's upward velocity is almost zero. At this time, the drop is not moving and the strong Marangoni flow mixes the liquid around the drop, thus greatly weakening the Marangoni flow itself, which will finally make the drop sink. During sinking, the drop first accelerates for a short period, then decelerates towards its density-matched position again. The drop repeats this bouncing cycle thereafter. For most of the time the drop is decelerating, so that the rising trajectory curves upwards and the sinking trajectory curves downwards. This is part of the reason why the trajectory is asymmetric. The other reason is that the average rising velocity of the drop is larger than the sinking velocity. For drops of higher viscosity, the rising velocity is smaller, thus their trajectories are less asymmetric, see figure \ref{fig:2}($a$-$c$).  \\
The bouncing cycles are different for drops of different radii, as can be seen from figure \ref{fig:2}($a$). They also depend on the degree of stratification and on the drop viscosities, see figure \ref{fig:2}($b,e$) and figure \ref{fig:2}($c,f$), respectively. We denote the highest position of the trajectory as $h_\mathrm{top}$ and the lowest position $h_\mathrm{bot}$, the rising time as $\tau_\mathrm{rise}$ and the sinking time $\tau_\mathrm{sink}$, see figure \ref{fig:2}($d$). The bouncing cycles can be characterized by these four quantities. 
In addition, figures \ref{fig:2}($g-i$) show the experimentally determined profiles of the mass density $\rho(y)$ and the interfacial surface tension $\sigma(y)$.  Approaching the uniform top layer gives rise to the pronounced non-linearity in the profile of the mass density, see figure \ref{fig:2}($h$).  Only small drops in very strong stratified liquids will reach this region, see the more detailed analysis in section \ref{sec:maxandminheight}. 
As mentioned above, the trajectories of the drops are highly asymmetric, where a fast rise is followed by a slow descent.  Although both oil viscosities show the same feature, this asymmetry is more dramatic for \SI{5}{cSt} than for \SI{100}{cSt} oil drops. 
In the following sections, we theoretically analyze how the four characteristic quantities of the bouncing cycle ($h_{\mathrm{top}}$, $h_{\mathrm{bot}}$, $\tau_{\mathrm{rise}}$, $\tau_{\mathrm{sink}}$) vary with the drop radius $R$, the strength of the stratification $\mathrm{d}w_\mathrm{e}/\mathrm{d}y$, and the drop viscosity $\mu^\prime$. 
The resulting scaling theory is then compared with the experimental results of the \SI{5}{cSt} drops in sections \ref{sec:maxandminheight} and \ref{sec:results5cst} and with the experimental and numerical results of the \SI{100}{cSt} drops in section \ref{sec:results100cst}.

\section{Dynamical analysis of the drop motion}
\label{sec:analysis}
The acceleration of the drop is caused by the forces acting on the drop: gravity and buoyancy $F_\mathrm{B}$, drag force $F_\mathrm{D}$ and a propulsion $F_\mathrm{M}$ caused by the Marangoni flow. Thus
\begin{equation}
m' \ddot{h} = F_{\mathrm{B}} + F_{\mathrm{D}} +  F_{\mathrm{M}},
\label{eq:eom}
\end{equation}
where $m'$ is the mass of the drop and $\ddot{h} = {\mathrm{d}^2h}/{\mathrm{d}t^2}$ the acceleration of the drop. The added mass force and the Basset history force are not taken into account because they are found to be negligible in the parameter space studied here \citep{yick2009enhanced}, see also section \ref{sec:results5cst}. They would at most modify prefactors. The Reynolds number is
\begin{equation}
\Rey=\frac{\vert\dot{h}\vert R}{\nu},
\end{equation}
where $\vert\dot{h}\vert$ is the absolute value of the drop velocity, $\nu=\mu/\rho$ is the kinematic viscosity of the mixture at the height of the drop, with $\mu$ and $\rho$ respectively the viscosity and density of the mixture at the height of the drop. 

In our case, the Reynolds number is small, see figure \ref{fig:5}, thus the drag force can be written as $F_\mathrm{D}=-\pi C^S_D \Rey \mu R\dot{h}/2$, where $C^S_D$ is the drag coefficient of the drop in the stratified liquid,
which will be discussed extensively in the following sections.
The buoyancy force is $F_\mathrm{B}=-V^\prime g(\rho^\prime-\rho)$ where $V^\prime$ is the volume of the drop. 
The propulsion is actually the viscous force caused by the Marangoni flow, $F_\mathrm{M}=k\mu V_\mathrm{M}R$, where $\mu$ is the viscosity of the mixture at the position of the drop, $V_\mathrm{M}$ is the Marangoni flow velocity and $k$ is a prefactor to be determined. 

Because the Marangoni flow is oscillatory for the bouncing drops \citep{li2021marangoni,li2022marangoni}, $V_\mathrm{M}$ is not constant.
When $V_\mathrm{M}$ is strong, the drop can reach its highest position $h_{\mathrm{top}}$ at which height the density of the ethanol-water mixture is $\rho_\mathrm{top}$; when $V_\mathrm{M}$ is weak, the drop will reach its lowest position $h_{\mathrm{bot}}$ at which height the density of the ethanol-water mixture is $\rho_\mathrm{bot}$. By balancing the Marangoni force with the buoyancy force i.e., when $F_{\mathrm{M}} = -F_{\mathrm{B}}$, we can obtain the density differences when the drop is at its highest \& lowest positions
\begin{equation}
\rho^\prime-\rho_\mathrm{top/bot}=\frac{k
\mu V_\mathrm{M}R}{V^\prime g}.
\label{eq:heq1}
\end{equation}
From \citet{young1959motion} we know that for the case of infinitely large diffusivity, zero density gradient and constant viscosity $\mu$, the Marangoni flow velocity at the equator of the drop is
\begin{equation}
V_{\mathrm{M}}\vert_\mathrm{equator} = -\frac{1}{2}\frac{\mathrm{d}\sigma}{\mathrm{d}w_{\mathrm{e}}}\frac{\mathrm{d}w_{\mathrm{e}}}{\mathrm{d}y}\frac{R}{\mu + \mu'},
\label{eq:marangonivelocity}
\end{equation}
where $\sigma$ is the interfacial tension between the drop and the mixture and $\mu'$ is the viscosity of the drop. But in our case, the diffusivity is not zero. Marangoni advection tends to smooth the concentration gradient close to the drop \citep{li2021marangoni}, thus weakening the Marangoni flow itself. The Marangoni flow at the highest position $V_\mathrm{M, top}$ and the lowest position $V_\mathrm{M, bot}$ can then be written as
\begin{equation}
V_{\mathrm{M, top}} \vert_\mathrm{equator}= -p\cdot \frac{1}{2}\frac{\mathrm{d}\sigma}{\mathrm{d}w_{\mathrm{e}}}\frac{\mathrm{d}w_{\mathrm{e}}}{\mathrm{d}y}\frac{R}{\mu + \mu'}, \quad\quad
V_{\mathrm{M, bot}}\vert_\mathrm{equator} = -q\cdot \frac{1}{2}\frac{\mathrm{d}\sigma}{\mathrm{d}w_{\mathrm{e}}}\frac{\mathrm{d}w_{\mathrm{e}}}{\mathrm{d}y}\frac{R}{\mu + \mu'}
\label{eq:Vtopbot}
\end{equation}
where $0<q<p<1$ are two prefactors to be determined. Since $p$ and $q$ represent the influence of Marangoni advection on the concentration field, we do not expect that they can be calculated beforehand. But we do expect them to vary with the viscosity of the drop. A higher drop viscosity normally leads to a weaker advection, thus the concentration field is less distorted by advection, so that the prefactor is larger. This trend has been confirmed by the levitating drops \citep{li2022marangoni}. According to \citet{young1959motion}, substituting Eq.(\ref{eq:Vtopbot}) and the volume of the drop into Eq.(\ref{eq:heq1}), we obtain
\begin{equation}
\rho^\prime-\rho_{\mathrm{top}} = -\alpha \cdot \frac{3}{2}\frac{\mu}{\mu + \mu'}\frac{\mathrm{d}\sigma}{\mathrm{d}y}\frac{1}{gR},\quad\quad \rho^\prime-\rho_{\mathrm{bot}} = -\beta \cdot \frac{3}{2}\frac{\mu}{\mu + \mu'}\frac{\mathrm{d}\sigma}{\mathrm{d}y}\frac{1}{gR}, 
\label{eq:rhotopbot}
\end{equation}
where $\alpha=kp$ and $\beta=kq$. Thus, $0<\beta<\alpha<k$ are two prefactors to be determined. Before we compare Eq.(\ref{eq:rhotopbot}) with the experimental values in section \ref{sec:maxandminheight}, we first analyse the governing time-scale of the bouncing intervals $\tau_\mathrm{rise}$ and $\tau_\mathrm{sink}$.

Because the Reynolds number in our experiments is very small, see figure \ref{fig:5}, the relevant time scale is effectively the time scale of the inertia-free system. That is to say, the acceleration occurs much faster than the force balance and is thus negligible. The time scale is then given by $F_\mathrm{B}+F_\mathrm{D}+F_\mathrm{M}=0$. 
For a linear gradient, $\rho=\rho^\prime+h\,\,\mathrm{d}\rho/\mathrm{d}y$, we thus have $F_\mathrm{B}=V^\prime gh\,\,\mathrm{d}\rho/\mathrm{d}y$. This yields
\begin{equation}
\dot{h} = \frac{b}{a} h + \frac{c}{a},
\label{eq:hdotp}
\end{equation}
where
\begin{equation}
\begin{aligned}
a &= \frac{\pi}{2} \mu R C^S_{D}\Rey, & b &=V' g \frac{d\rho}{dy}, & c &= k\mu V_{\mathrm{M}}.
\end{aligned}
\label{eq:definitions}
\end{equation}

Not only the drag coefficient $C^S_{D}$ but also the Reynolds number $\Rey$ depends on the position of the drop. However, according to \citet{zvirin1975settling}, Eq.(\ref{eq:hdotp}) can have an accuracy to first order when evaluating the drag coefficient $C^S_{D}$ and the Reynolds number $\Rey$ at the instant the drop reaches its peak velocity $\dot{h}_\mathrm{p}$. Eq.(\ref{eq:hdotp}) implies an exponential behaviour \citep{zvirin1975settling,li2019bouncing} with a governing time-scale for the drop to reach its equilibrium position through either rising or sinking, which is
\begin{equation}
\tau_\mathrm{rise/sink}\sim\tau_1\approx -\frac{a}{b}\sim\frac{\nu_{\mathrm{p}} C^S_{D,\mathrm{p}}\Rey_\mathrm{p}}{N^2_{\mathrm{p}}R^2}
\label{eq:tau}
\end{equation}
where
\begin{equation}
N_{\mathrm{p}}=\sqrt{-\frac{g}{\rho}\frac{\mathrm{d}\rho}{\mathrm{d}y}}
\end{equation}
is the Brunt-V\"a\.is\"al\"a frequency. Note that Eq.(\ref{eq:tau}) is consistent with the results of \citet{zvirin1975settling}. To evaluate the quantities in Eq.(\ref{eq:tau}), ethanol weight fractions at the positions where the drops reach their peak velocities (during rising and sinking) are used to obtain the corresponding density $\rho_{\mathrm{p}}$, viscosity $\mu_{\mathrm{p}}$ and interfacial tension $\sigma_{\mathrm{p}}$ (see Appendix \ref{appA} for the concentration dependence of these properties).  
It has become apparent that the drag coefficient $C_D^S$ is of great importance for the overall dynamics. Therefore, we provide more details regarding the drag on spherical objects in stratified liquids in section \ref{sec:dragexplain}, followed by a discussion on the experimental and numerical results in sections \ref{sec:results5cst}\&\ref{sec:results100cst}. 

\begin{figure}
  \centerline{\includegraphics[width=0.9\textwidth]{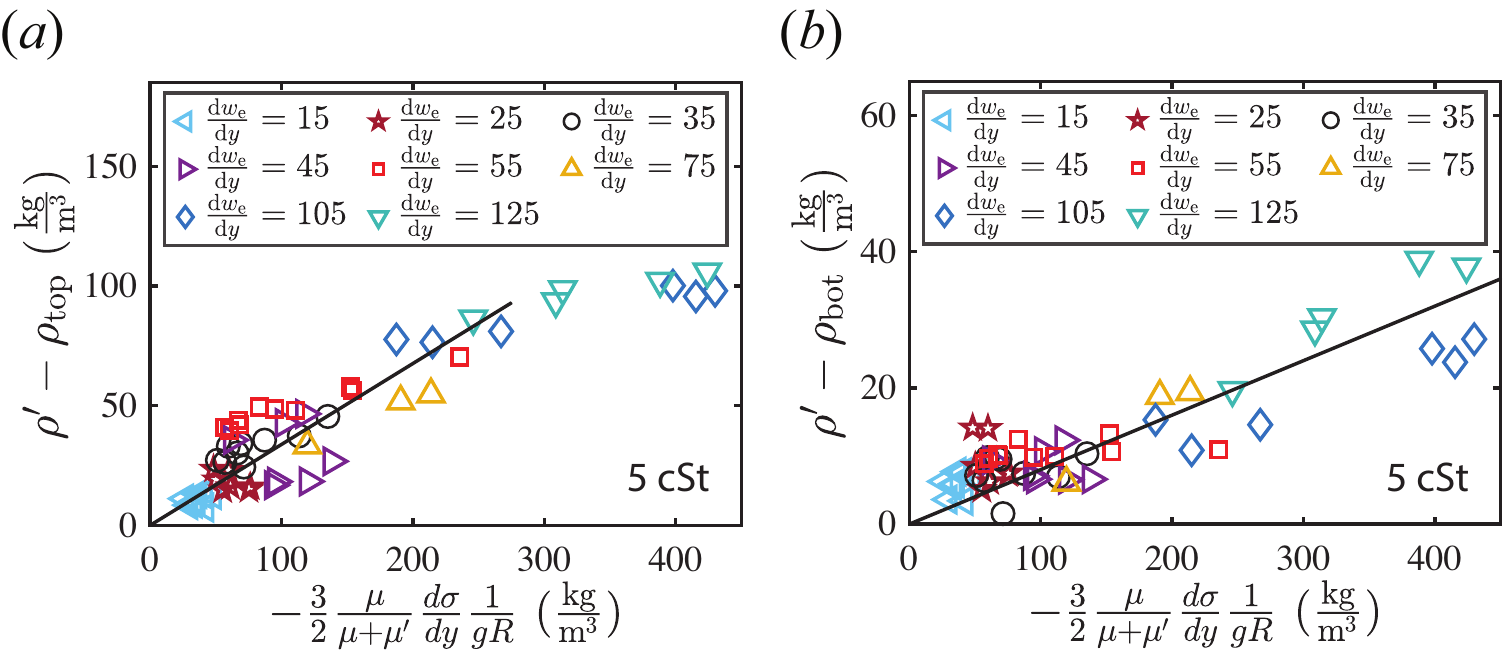}}
  \caption{($a$) Maximum and ($b$) minimum bouncing height of \SI{5}{cSt} silicone oil drops of different radii in linearly stratified ethanol-water mixtures with indicated stratification strengths in m$^{-1}$.  The deviations from the linear trend at larger weight fraction gradients can be explained by the occurrence of ceiling effects, i.e., the drop approaching the upper region of constant density. The measured prefactors are $\alpha_{\mathrm{5cSt}} = 0.33$ and $\beta_{\mathrm{5cSt}} = 0.08$. } 
\label{fig:3}
\end{figure}

\section{The density differences at the maximum and minimum bouncing positions}
\label{sec:maxandminheight}
\begin{figure}
  \centerline{\includegraphics[width=0.9\textwidth]{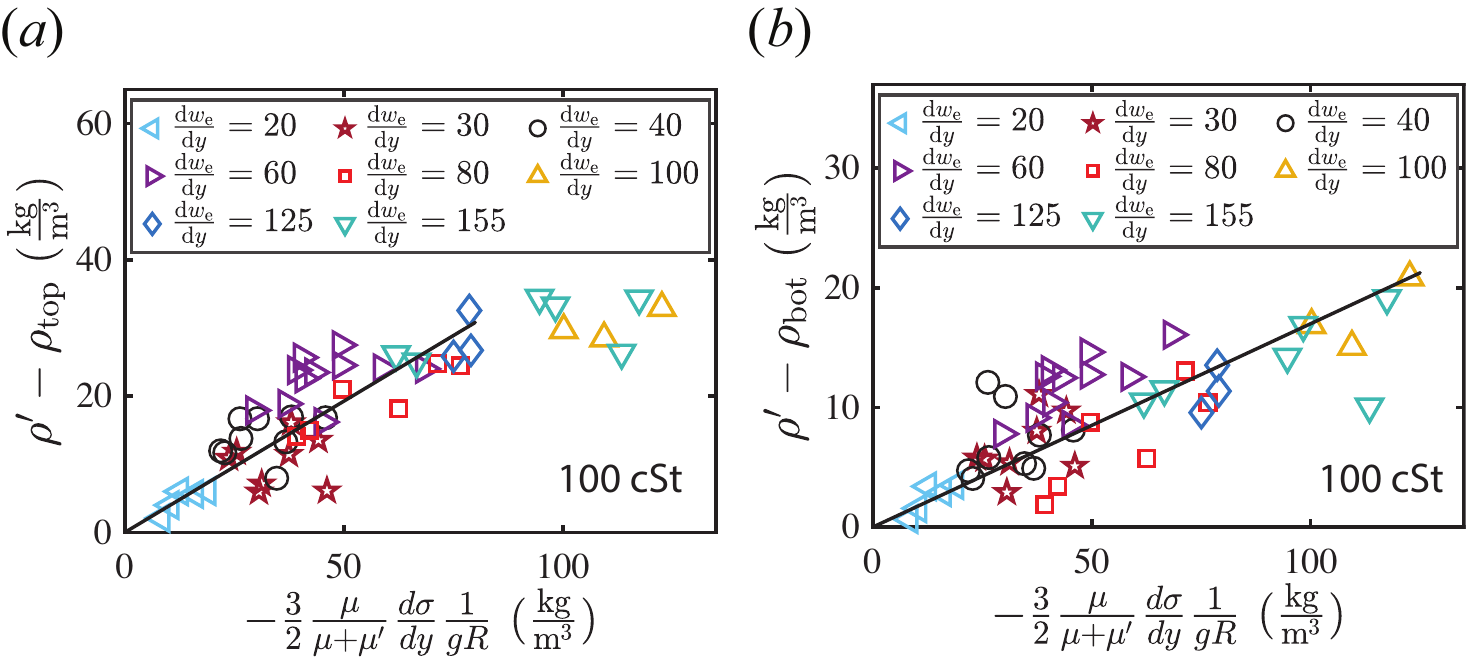}}
  \caption{($a$) Maximum and ($b$) minimum bouncing position of \SI{100}{cSt} silicone oil drops of different radii in linearly stratified ethanol-water mixtures with indicated stratification strengths in m$^{-1}$.  The deviations from the linear trend at larger weight fraction gradients can be explained by the occurrence of ceiling effects i.e., the drop approaching the upper region of the constant density. The measured prefactors are $\alpha_{\mathrm{100cSt}}= 0.39$ and $\beta_{\mathrm{100cSt}}= 0.20$.}
\label{fig:4}
\end{figure}

In this section we compare the theoretically predicted density differences at the maximum and minimum bouncing positions with the experimentally measured ones.  
The quantities on the right hand sides of Eq.(\ref{eq:rhotopbot}) for \SI{5}{cSt} drops are calculated (excluding the prefactors $\alpha$ and $\beta$) and plotted against the experimentally measured density differences $\rho^\prime-\rho_\mathrm{top}$ and $\rho^\prime-\rho_\mathrm{bot}$ in figure \ref{fig:3}($a$) and (b), respectively. 
The position dependent material properties $\mu$ and $\sigma$ are evaluated at either the highest or lowest position, correspondingly.
The results for \SI{100}{cSt} are shown in figure \ref{fig:4}. 
 
In both cases the density differences follow a linear trend initially, as predicted by Eq.(\ref{eq:rhotopbot}), but are accompanied by a considerable amount of scatter. The scatter originates from the non-linear profiles of both the viscosity $\mu$ and surface tension $\sigma$ as functions of the ethanol weight fraction $w_{\mathrm{e}}$, see Appendix \ref{appA}. 

From this procedure the prefactors $\alpha$ and $\beta$ can be determined and are found to be $\alpha_{\SI{5}{cSt}}=0.33$, $\beta_{\SI{5}{cSt}}=0.08$, $\alpha_{\SI{100}{cSt}}=0.39$ and $\beta_{\SI{100}{cSt}}=0.20$,  respectively.  As drops move higher, the density differences, or equivalently the bouncing heights saturate, because the drops are reaching the top uniform layer (see figure \ref{fig:1}), which forms the "ceiling" of the bouncing trajectory.  Not surprisingly, only small drops in large concentration gradients can reach the ceiling. 

Comparing the obtained prefactors, it appears that they depend on the viscosity of the oil. To rationalize this observation we recall the origin of the prefactor $3/2$ \citep{young1959motion} on the right hand sides of Eq.(\ref{eq:rhotopbot}), where an infinitely large diffusivity was assumed. Given the fact that the Marangoni advection is important in our case, this assumption no longer holds, causing a reduction of this prefactor, hence $0<\beta<\alpha<1$.  And since the Marangoni flow is inversely proportional to the viscosity of the oil $\mu'$, see Eq. (\ref{eq:marangonivelocity}),  $\alpha$ and $\beta$ will increase with $\mu'$. 
Finding the exact functional form of $\alpha$ and $\beta$ on $\mu'$ is beyond the scope of the present paper. 

\section{Drag on spherical objects in stratified media}
\label{sec:dragexplain}

As discussed in section \ref{sec:analysis}, the governing time-scale of the motion of the drop (Eq. (\ref{eq:tau})) depends on the drag coefficient of the drops in the stratified liquid $C_D^S$. As mentioned in the Introduction, there are only a few studies on the topic of a drop moving inside stratified liquids with the presence of Marangoni flow. In particular, there are no available data on the drag coefficient of a drop moving inside such stratifications. However, we think the drag coefficient of a solid sphere in such stratifications could be used here, since we are only interested in the scaling of the rising \& sinking times and the drag coefficient of drops and solid spheres normally only differ by a prefactor \citep{hadamard1911mouvement, rybczynski1911uber}. This is confirmed later in section \ref{sec:results5cst}. In this section, we will therefore provide a brief overview on the current understanding of drag on solid spheres in stratified liquids. 

Analytically, the settling of a spherical body in a stratified liquid has been studied in the past by several authors. The relevant parameters for this problem are the Froude number, defined as
\begin{equation}
\Frou = \frac{\vert \dot{h} \vert}{NR},
\label{eq:Froude}
\end{equation}
the Péclet number, defined as
\begin{equation}
\Pen = \frac{\vert \dot{h} \vert R}{D},
\label{eq:Peclet}
\end{equation}
and the Richardson number, defined as
\begin{equation}
\Rich = \frac{\Rey}{\Frou^2} = \frac{N^2R^3}{\nu \vert \dot{h} \vert},
\end{equation}
where $D$ is the solute diffusivity.
For example, \citet{zvirin1975settling} derived that in the limit of $\Rey\ll 1$ and $\Frou\ll 1$ and under the assumption that advection dominates, i.e.,  $\Pen \rightarrow  \infty$, the drag coefficient scales as
\begin{equation}
C_D^S \sim \Rich^{1/3}/\Rey. 
\end{equation}

\begin{figure}
\centerline{\includegraphics[width=0.55\textwidth]{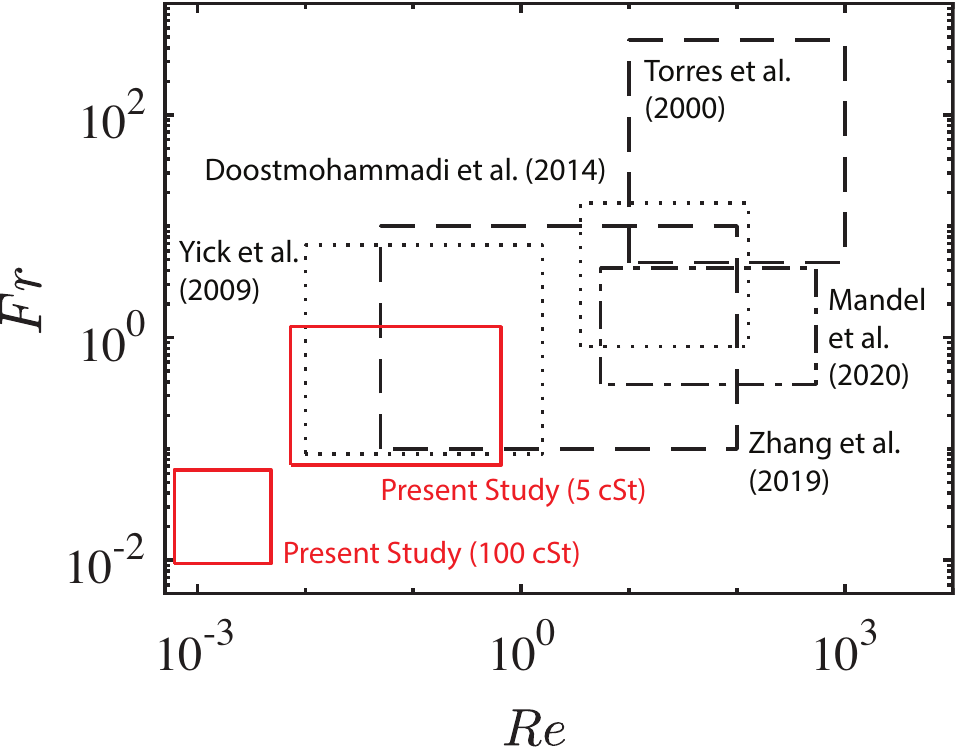}}
\caption{Black boxes are parameter spaces studied previously on the settling of spherical objects in linearly stratified liquids \citep{torres2000flow, yick2009enhanced,  doostmohammadi2014numerical, zhang2019core, mandel2020retention}. Both numerical and experimental results are included. The red boxes are parameter spaces of the present study in terms of $\Rey_\mathrm{p}$ and $\Frou_\mathrm{p}$, for \SI{5}{cSt} and \SI{100}{cSt} silicone oil drops, respectively. The limit of $\Rey\ll 1$ and $\Frou\ll 1$ was studied analytically by \citet{zvirin1975settling} for $\Pen \rightarrow \infty$ and by \citet{candelier2014history} for the opposite limit $\Pen \ll 1$.}
\label{fig:5}
\end{figure}

In the opposite limit, when diffusion dominates advection, i.e., when $\Pen \ll 1$, \citet{candelier2014history} derived that 
\begin{equation}
C_D^S \sim (\Pen \Rich)^{1/4}/\Rey.
\end{equation}
It has recently been shown by \citet{mehaddi2018intertial} that both of these scaling relations can be obtained from the very same derivation, where the expressions above are simply limiting cases.  

The drag coefficient of a particle settling in a density stratification has also been investigated experimentally and numerically \citep{srdic1999gravitational,torres2000flow, yick2009enhanced,zhang2019core},  and different forms of the drag coefficient have been suggested in different parameter ranges. Some of the studied parameter ranges in terms of Froude number versus Reynolds number are summarized in figure \ref{fig:5}. 
The figure also shows the parameter range of the present study, see the two red boxes for \SI{5}{cSt} and \SI{100}{cSt} drops, respectively. It can be seen that the parameter range of the \SI{5}{cSt} drops overlaps almost entirely with that of \citet{yick2009enhanced}, where the drag coefficient was empirically determined as
\begin{equation}
C_D^S \sim \frac{\Rich^{0.51 \pm 0.11}}{\Rey}.
\label{eq:CDYick}
\end{equation}
An argument for the discrepancy between this result and the analytical solutions was provided by \citet{zhang2019core}. In their work they defined three different stratification regimes depending on the relative magnitudes of three length scales: the viscosity length scale $l_{\nu}$, the diffusivity length scale $l_{D}$, and the stratification length scale $l_s$. If $l_s \ll l_{D} \ll l_{\nu}$ (Regime 1), then the drag coefficient scales as predicted by \citet{candelier2014history}. Otherwise, if $l_{D} \ll l_{s} \ll l_{\nu}$ (Regime 2), then \citet{zvirin1975settling} gave the correct prediction. If $l_{D} \ll l_{\nu} \ll l_{s}$  (Regime 3), then $C_D^S \sim (\Frou \Rey)^{-1}$ \citep{zhang2019core}. They argue that, since the experimental and numerical results by \citet{yick2009enhanced} mix two of these asymptotic regimes, the obtained scaling relation is an \textit{ad hoc} approximation.  

Since we do not know the appropriate scaling relation of the drag coefficient \textit{a priori}, especially for the \SI{100}{cSt} oil drops that fall in a parameter regime not yet studied in the available literature (see figure \ref{fig:5}), we assume a general expression as 
\begin{equation}
C_D^S \sim \Rich^q / \Rey, 
\end{equation}
where $q$ is an exponent to be determined. The scaling relation of the dominant time scale, Eq.(\ref{eq:tau}), can then be rewritten as:
\begin{equation}
\tau_{\mathrm{rise/sink}} \sim \frac{R}{\vert \dot{h} \vert_{\mathrm{p,rise}/\mathrm{sink}}}\Rich^{(q-1)}_\mathrm{p}.
\label{eq:taunew}
\end{equation}
In the following section Eq. (\ref{eq:taunew}) will be evaluated using the experimentally determined time intervals and drop velocities for \SI{5}{cSt} and \SI{100}{cSt} oil drops, respectively. The obtained values of $q$ are then compared to the above mentioned scaling relations (section \ref{sec:results5cst}) and to our numerical simulations (section \ref{sec:results100cst}). 

In the following sections we first discuss the experimental results for the low viscosity drops, then the experimental and numerical results for the high viscosity drops. 

\section{Results of the low viscosity drops: \SI{5}{cSt}}
\label{sec:results5cst}

\begin{figure}
  \centerline{\includegraphics[width=0.85\textwidth]{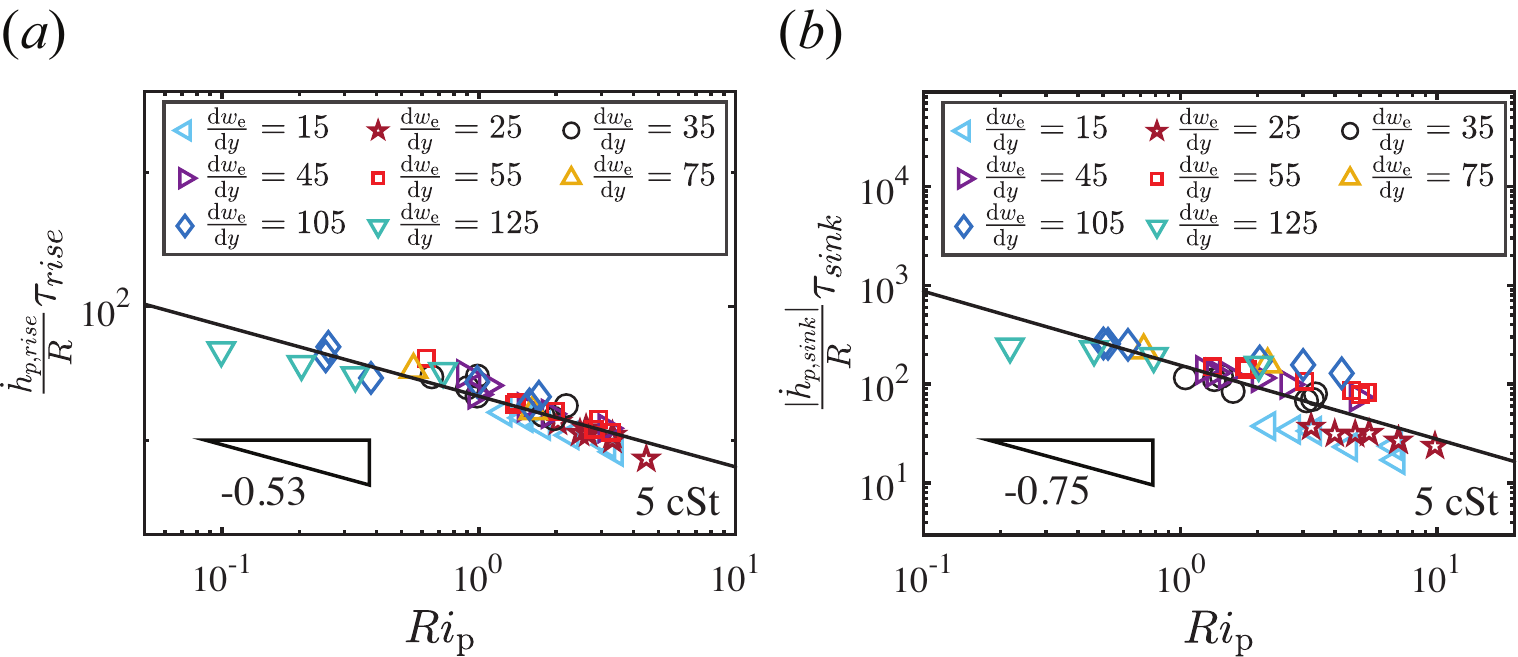}}
  \caption{Non-dimensionalised rising ($a$) and sinking ($b$) times of \SI{5}{cSt} silicone oil drops of different radii in linearly stratified ethanol-water mixtures with indicated stratification strengths as functions of $\Rich_{\mathrm{p}}$ on a double logarithmic scale. The solid line shows the best fit through the experimental data.}
\label{fig:6}
\end{figure}

The result after evaluating Eq. (\ref{eq:taunew}) with the experimentally determined time intervals and drop velocities for \SI{5}{cSt} oil drops during rising and sinking are shown in figure~\ref{fig:6}($a$)\&($b$), respectively. The solid lines show the best fits through the experimental data. It follows that, since $q-1 = -0.53$ for rising and $q-1 = -0.75$ for sinking, a scaling relation of the drag coefficient during rising and sinking is obtained as
\begin{equation}
\begin{aligned}
C^S_{D,\mathrm{rise,\SI{5}{cSt}}} &\sim \frac{\Rich^{0.47}_{\mathrm{p}}}{\Rey_{\mathrm{p}}}, & C^S_{D,\mathrm{sink,\SI{5}{cSt}}} &\sim \frac{\Rich^{0.25}_{\mathrm{p}}}{\Rey_{\mathrm{p}}}.
\end{aligned}
\end{equation}
The scaling relations are clearly different. For the rising drop the determined scaling relation is in good agreement with the empirical result  of \citet{yick2009enhanced}, which we write as $C_D^S\sim \Rich^{0.51\pm 0.11}/Re$, see Eq.(\ref{eq:CDYick}).
Although the latter was established for a solid particle, here it is shown that the same relation also applies to drops. 
The Marangoni velocity $V_{\mathrm{M}}$ does not influence the governing time-scale for rising or sinking (see Eq.(\ref{eq:hdotp})) and will influence the drag coefficient $C_D^S$ only by changing the parameter range (\Frou, \Rey).
Thus, as long as the parameter range fits, the results of the drag coefficient $C_D^S$ on particles can be directly applied to drops, which is in line with the observation that the studies overlap almost entirely in the parameter space shown in figure \ref{fig:5}. 

For the sinking drop these scaling relations do not match. The obtained result seems to be in agreement with the scaling relation as predicted by \citet{candelier2014history} $C_D^S \sim (\Pen \Rich)^{1/4}/\Rey$, which is the asymptotic limit of the viscous-diffusive regime where diffusion dominates over advection. This is because, as discussed in section \ref{sec:characteristics},  during sinking, the drop is surrounded by an entrained shell where diffusion is dominant \citep{li2019bouncing}.  
The fact that $C^S_{D,\mathrm{rise}} > C^S_{D,\mathrm{sink}}$ is the second reason mentioned earlier why the bouncing trajectory is asymmetric.

\section{Results of the high viscosity drops: \SI{100}{cSt}}
\label{sec:results100cst}

Before evaluating Eq. (\ref{eq:taunew}) with the experimentally measured values it is important to realize that the parameter space spanned  by the high viscous oil is not covered by any available literature, see figure \ref{fig:5}.  Although analytical results do exist for limiting cases, as discussed in section \ref{sec:dragexplain}, they as well as other similar studies \citep{lee2019sedimentation,  dandekar2020motion, shaik2020drag} relied on the crucial assumption $\Rich \ll 1$, i.e., weakly stratified liquids, due to the applied mathematical method of asymptotic matching.  Since in our case for the \SI{100}{cSt} drops $1.7 \leq \Rich_{\mathrm{p}} \leq 27.6$, the liquid is strongly stratified and it is questionable whether the above scaling relations apply to our present study.  A new theoretical derivation of the drag coefficient of drops settling in a strongly stratified liquid including Marangoni effects is beyond the scope of the present work.  
Thus, numerical simulations covering the same parameter range are performed to provide independent verification. 
Details of the numerical simulations have been provided in section \ref{NumericalMethods}. Here we discuss the numerical results and compare them with the experiments.

\subsection{Numerical results}

\begin{figure}
  \centerline{\includegraphics[width=0.95\textwidth]{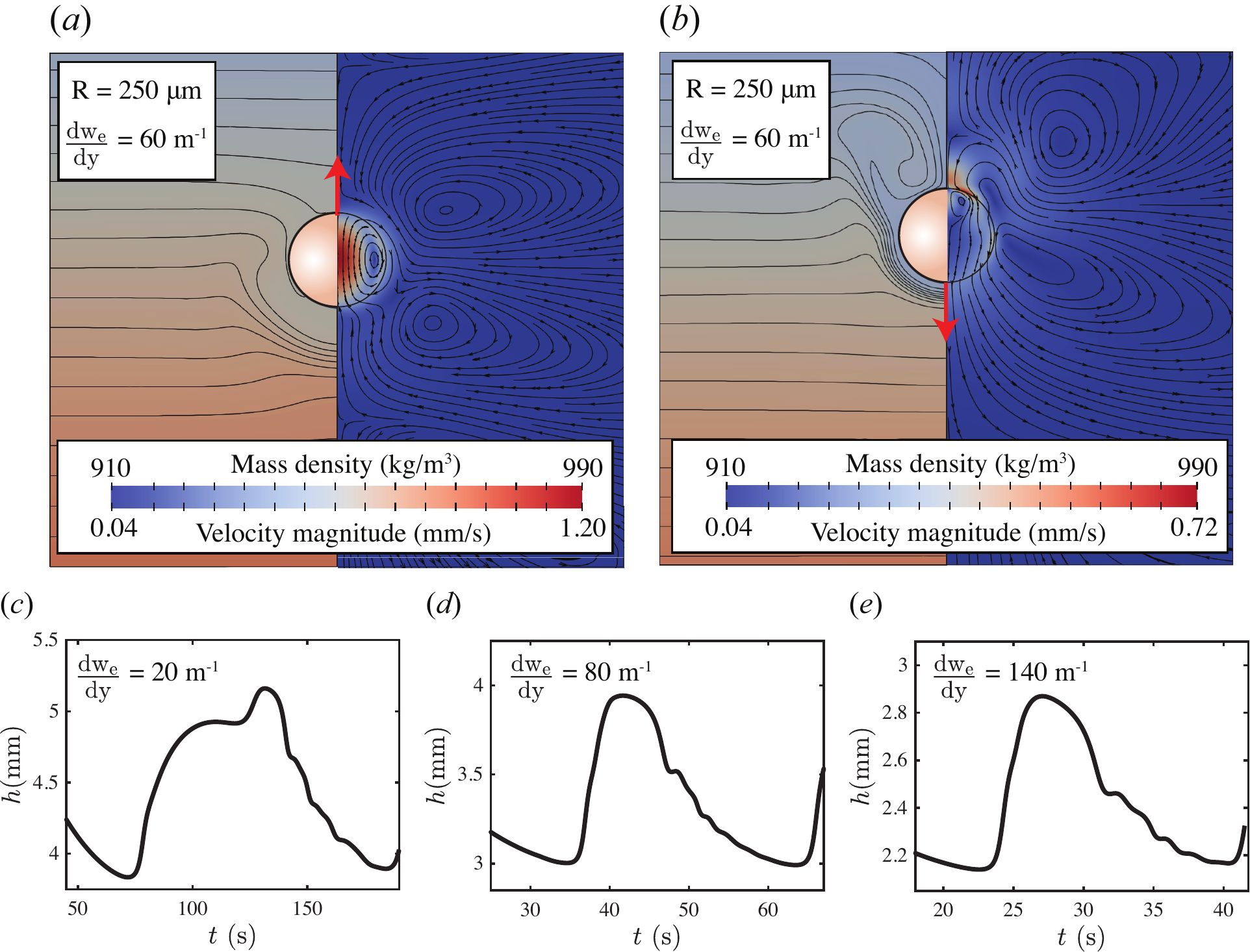}}
  \caption{Snapshot of the numerical simulation showing the mass density of the mixture (left) and the velocity magnitude (right) around a ($a$) rising and ($b$) sinking \SI{100}{cSt} oil drop as it reaches its peak velocity. ($c$)-($e$) Temporal evolution of the drop's height $h(t)$ with $R = \SI{250}{\micro \meter}$ for three different stratifications. }
\label{fig:7}
\end{figure}

The numerical simulations are initialized by placing a \SI{100}{cSt} drop ($\SI{250}{\micro \meter} \leq R \leq \SI{400}{\micro \meter}$) above its density matches position, i.e., $h = 0$, into a linearly stratified ethanol-water mixture ($\SI{20}{\per \meter} \leq \mathrm{d}w_{\mathrm{e}}/\mathrm{d}y \leq \SI{140}{\per \meter}$), with corresponding uniform top and bottom layers.  As the drop descends towards this equilibrium position, at a certain moment in time Marangoni effects start to become dominant and the bouncing is initiated, the same as in the experiments.  Since the simulations demand a rather small time step due to the high Péclet number and thus considerable CPU time, only the first complete bouncing cycle is considered when analysing the numerical results. This is in contrast with the experiments, where the third bouncing cycle is used. 

Typical snapshots of the numerical simulation as the drop reaches its peak velocity during rising and sinking are shown in figures \ref{fig:7}($a$)\&($b$), respectively.  We show the mass density of the mixture and how mixing occurs in close proximity to the drop, as well as the velocity magnitude inside and outside the drop. In the latter clear vortical structures arise that are induced by the baroclinic torque following the deflection of the isopycnals.  As the drop rises, a strong Marangoni flow is visible at the interface causing strong internal circulations in the drop,  see figures \ref{fig:7}($a$).  The isopycnals are compressed below the drop as lighter fluid is advected downwards by the Marangoni flow. During sinking,  the high velocity region has shifted towards the apex of the drop and the strong internal circulation in the drop has stopped, see figure \ref{fig:7}($b$).  
This leads to the conclusion that during sinking Marangoni effects are indeed rendered ineffective, see Appendix \ref{appC}.
The deflection of the isopycnals has become more significant as the drop is dragging liquid down.  

In addition, the trajectories of three bouncing drops with the same size but in different stratifications are plotted in figures \ref{fig:7}($c-e$).  Qualitatively the numerically simulated bouncing cycles show great resemblance to the experimental observations, see figure \ref{fig:2}.  A quick rise is followed by a slower descent, where the asymmetry in the bouncing cycle is less dramatic as for high viscosity oils.  
Also, the bouncing period seems to shorten with stronger stratifications and the bouncing amplitude becomes smaller, the same as in the experiments.
The second rise in figure \ref{fig:7}($c$) for a small drop in a relatively weakly stratified ethanol-water mixture has also been observed experimentally, see figure \ref{fig:2}($d$), although less pronounced. 

\subsection{Experimental and numerical comparison}

\begin{figure}
  \centerline{\includegraphics[width=\textwidth]{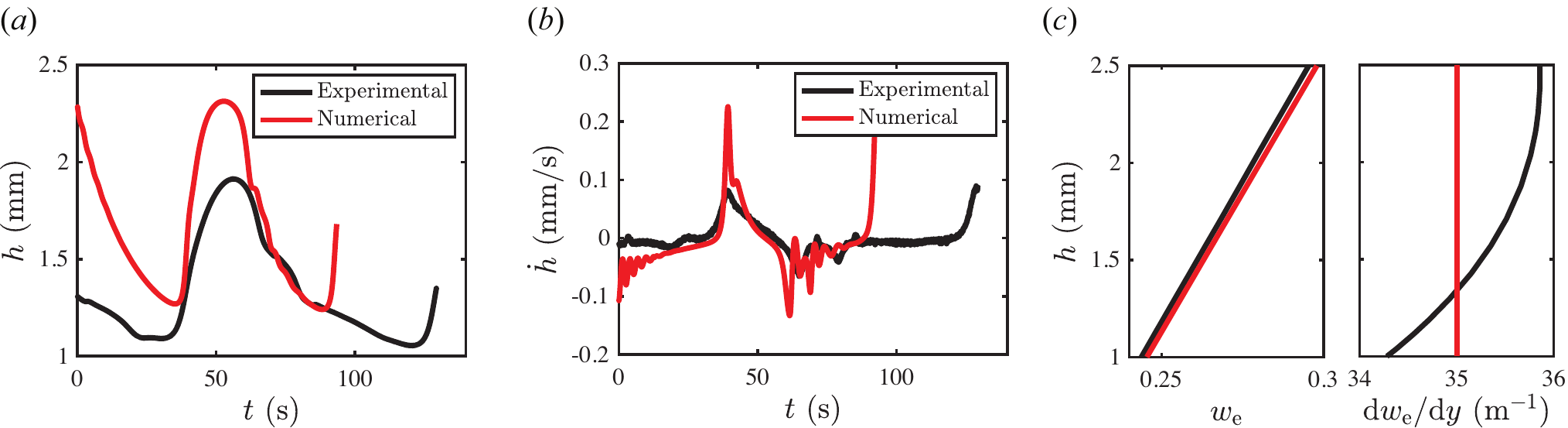}}
  \caption{Experimentally measured (black) and numerically determined (red) ($a$) trajectory $h(t)$ and ($b$) velocity $\dot{h}(t)$ of a \SI{100}{cSt} oil drop with $R = \SI{280}{\micro \meter}$ and $\mathrm{d}w_{\mathrm{e}}/\mathrm{d}y = \SI{35}{\per \meter}$ (See Supplementary Movie 1 for more details).  ($c$) The background concentration profiles and corresponding gradients. }
\label{fig:8}
\end{figure}

Now we are in a position that allows us to compare the experimental and numerical results one-on-one. For this specific case a \SI{100}{cSt} oil drop with $R = \SI{280}{\micro \meter}$ is placed inside a stratified ethanol-water mixture where $\mathrm{d}w_{\mathrm{e}}/\mathrm{d}y = \SI{35}{\per \meter}$.  Figures \ref{fig:8}($a$)\&($b$) compare the drop's trajectories and velocity profiles, respectively.  Qualitatively, again,  the agreement is good. Quantitatively, there are some differences.
For example, the overall bouncing period is shorter in the simulations ($T = \SI{52.3}{\second}$) compared to the experiments ($T = \SI{88.8}{\second}$) and the peak velocities reached in the numerics ($\dot{h}_{\mathrm{p,rise,num}} = \SI{0.24}{\milli \meter \per \second}$ and $\dot{h}_{\mathrm{p,sink,num}} = \SI{-0.13}{\milli \meter \per \second}$) exceed those of the experiment ($\dot{h}_{\mathrm{p,rise,exp}} = \SI{0.08}{\milli \meter \per \second}$ and $\dot{h}_{\mathrm{p,sink,exp}} = \SI{-0.06}{\milli \meter \per \second}$). In addition, the top and bottom positions of the bouncing curve from the simulation ($h_{\mathrm{top,num}} = \SI{2.3}{\milli \meter}$ and $h_{\mathrm{bot,num}} = \SI{1.3}{\milli \meter}$) are shifted slightly above the experimentally determined ones ($h_{\mathrm{top,exp}} = \SI{1.9}{\milli \meter}$ and $h_{\mathrm{bot,exp}} = \SI{1.1}{\milli \meter}$).
Both discrepancies might be explained by the fact that the Marangoni force, caused by Marangoni advection at the interface, is overestimated in the simulations. This would give rise to a quicker ascent, causing the drop to reach greater heights and thus reaching larger velocities during its descent, as the drop is further away from its density matched position.
Comparing the background concentration profiles and the corresponding concentration gradients in figure \ref{fig:8}($c$) shows that although the concentration profiles look very similar, the local concentration gradient varies in the experiment, whereas it remains constant in the numerical simulation. 
Consequently, the concentration gradient $\mathrm{d}w_{\mathrm{e}}/\mathrm{d}y$ the drop "feels" at $h_{\mathrm{bot}}$ is smaller in the experiments than in the numerics.
Since $V_{\mathrm{M}} \sim \mathrm{d}w_{\mathrm{e}}/\mathrm{d}y$, see Eq.\ref{eq:marangonivelocity}, Marangoni advection will be slightly larger in the numerical simulation.    

\begin{figure}
  \centerline{\includegraphics[width=0.85\textwidth]{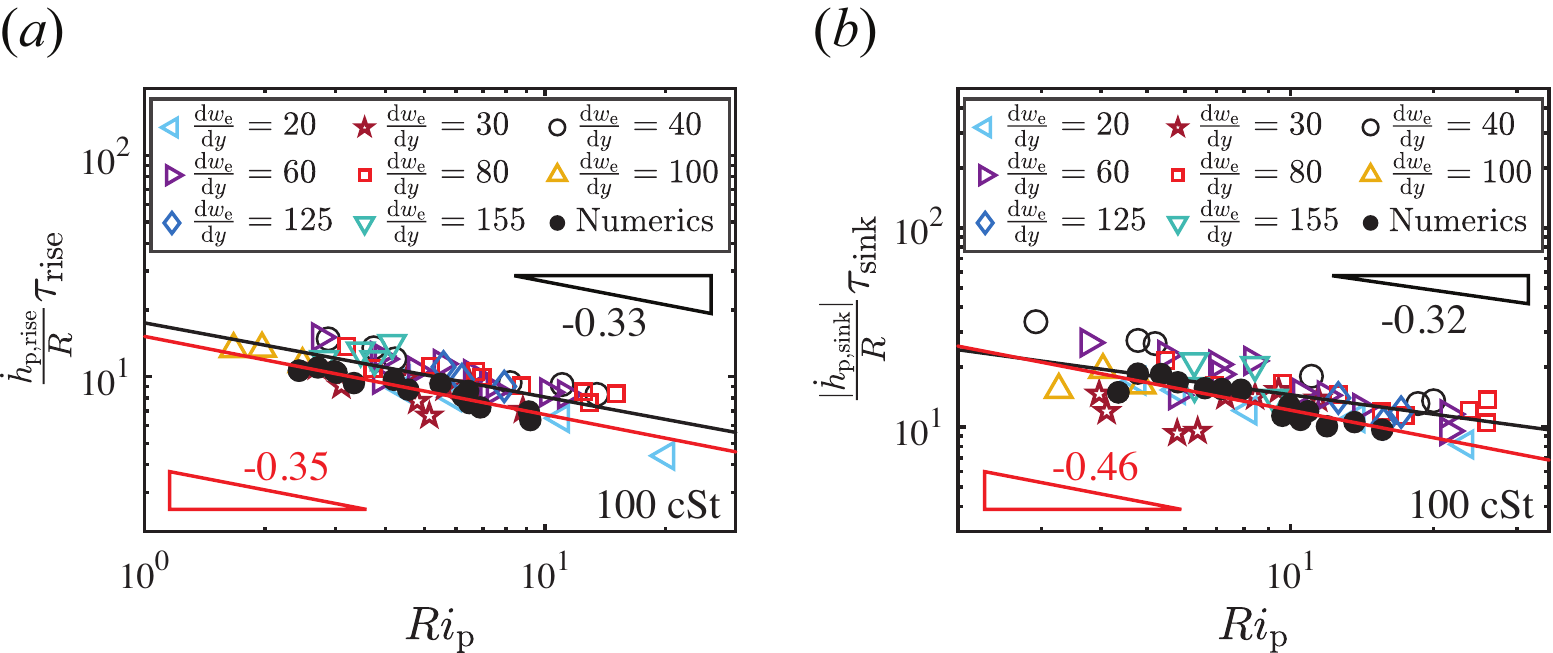}}
  \caption{Non-dimensionalised rising ($a$) and sinking ($b$) times of \SI{100}{cSt} silicone oil drops of different radii in linearly stratified ethanol-water mixtures with indicated stratification strengths as functions of $\Rich_{\mathrm{p}}$ on a double logarithmic scale. The black (red) line shows the best fit through the experimental (numerical) data.}
\label{fig:9}
\end{figure}

Finally,  Eq. (\ref{eq:taunew}) is evaluated with the experimentally and numerically determined time intervals and drop velocities for \SI{100}{cSt} oil drops during rising and sinking.  As in section \ref{sec:results5cst} the aim is to determine the exponent $q$ to obtain the scaling relation of the drag coefficient $C_D^S \sim \Rich^q / \Rey$.  In figure \ref{fig:9}($a$)\&($b$) the results are shown for both rising and sinking, respectively. The empty symbols represent the experimental results and filled circles represent the numerical results. 

Based on the best fit through the data, indicated by the solid lines, it holds for the experiments that
\begin{equation}
\begin{aligned}
C^{S,\mathrm{exp}}_{D,\mathrm{rise,\SI{100}{cSt}}} &\sim \frac{\Rich^{0.67}_{\mathrm{p}}}{\Rey_{\mathrm{p}}}, & C^{S,\mathrm{exp}}_{D,\mathrm{sink,\SI{100}{cSt}}} &\sim \frac{\Rich^{0.68}_{\mathrm{p}}}{\Rey_{\mathrm{p}}},
\end{aligned}
\end{equation}
and for the simulations that
\begin{equation}
\begin{aligned}
C^{S,\mathrm{num}}_{D,\mathrm{rise,\SI{100}{cSt}}} &\sim \frac{\Rich^{0.65}_{\mathrm{p}}}{\Rey_{\mathrm{p}}}, & C^{S,\mathrm{num}}_{D,\mathrm{sink,\SI{100}{cSt}}} &\sim \frac{\Rich^{0.54}_{\mathrm{p}}}{\Rey_{\mathrm{p}}}.
\end{aligned}
\end{equation}
Considering these results our study suggest that the drag coefficient of the high viscosity drops in strongly stratified liquids scales as
\begin{equation}
\begin{aligned}
C^S_{D,\mathrm{rise,\SI{100}{cSt}}} &\sim \frac{\Rich^{0.66 \pm 0.01}}{\Rey}, & C^S_{D,\mathrm{sink,\SI{100}{cSt}}} &\sim \frac{\Rich^{0.61 \pm 0.07}}{\Rey}.
\end{aligned}
\label{eq:scaling100cst}
\end{equation}
It is found that the scaling relations differ from the analytical predictions by \citet{zvirin1975settling} and \citet{candelier2014history}, as well as from the empirical relation obtained by \citet{yick2009enhanced}. 

An argument for this discrepancy has already been addressed in section \ref{sec:dragexplain}. There, three different stratification regimes are introduced, depending on the relative magnitudes of the viscosity length scale $l_{\nu}$, the diffusivity length scale $l_{\mathrm{\kappa}}$, and the stratification length scale $l_s$ \citep{zhang2019core}.  As mentioned, the experimental and numerical results by \citet{yick2009enhanced} mix two of these asymptotic regimes and the obtained scaling relation is therefore an \textit{ad hoc} approximation.  Provided that $\Rey \ll 1$ and $\Pen \gg 1$, where $\Pen = \Rey \Pran$ with $\Pran = \nu/D$,  the domain of existence for each regime respectively is \citep{zhang2019core}: 
\begin{equation}
\Frou \ll \Rey^{1/3} \Pran^{-1/6}, \quad \quad \Pran^{-1/2} \ll \Frou \ll \Rey^{-1}, \quad \quad \Frou \gg \Rey^{-1}.
\end{equation}
Evaluating these conditions, it becomes evident that in the present study of the \SI{100}{cSt} drops the conditions of Regimes 1 and 2 are met simultaneously and that a mix of these regimes occurs; hence the deviation from the theoretical predictions by \citet{zvirin1975settling}. 

\section{Conclusion and outlook}\label{sec:conclusion}
To summarize, the bouncing dynamics of drops of different viscosities in stably stratified liquids with the presence of Marangoni flow is studied theoretically, experimentally and numerically. The main characteristics of the bouncing cycle have been discussed and the importance of the findings by \citet{young1959motion} regarding the prediction of the maximum and minimum bouncing positions has become evident.  
Based on our derivation of the scaling relation of the governing time scale in which the drop reaches its equilibrium position through either rising or sinking, it has become apparent that the scaling relation of the drag coefficient in the stratified liquid $C_D^S$ is of great importance.
To this end, the experimentally determined quantities of the rising and sinking times, as well as the peak velocities reached during rising and sinking are used to obtain the appropriate scaling relation of $C_D^S$ for drops in strongly stratified liquids.
For low viscous oil drops it was found that the drag coefficient follows the scaling relations obtained from literature. The significant difference between the relations obtained for rising and sinking explains the high asymmetry of the bouncing cycle. 
For the high viscosity oil drops the scaling relation of the drag coefficient is not available in the literature. Thus, to seek independent verification, numerical simulations are performed mimicking the experiments for $1.7 \leq \Rich_{\mathrm{p}} \leq 27.6$. 
This also allowed for a one-to-one comparison between experiments and numerics, where it has been found that qualitatively the agreement between the bouncing cycles is good. 
For both results, experimental and numerical, scaling relations are obtained for the high viscosity oil drops during rising and sinking in strongly stratified liquids.

It is also found that when in the same parameter range, the scaling of the drag coefficient of a solid sphere could be applied to that of a drop, with or without Marangoni flow. This is supported by the results for \SI{5}{cSt} drops. Thus, the extensive knowledge on drag coefficients of solid spheres in stratified media can be of help to that of drops, which has rarely been explored. We also found that in the parameter space of the \SI{100}{cSt} drops, the drag coefficient under stratified conditions has a scaling $C_D^S\sim \Rich^{0.66\pm0.01}/\Rey$.

As our work has shown, new insight has still to be discovered on the rising and sinking  drops and bubbles in strongly stratified liquids.  In particular,  the dominant mechanism behind the drag enhancement, only recently discovered by \citet{zhang2019core}, shows the complexity and richness of such hydrodynamical systems.  Whereas the drag coefficient of spherical objects in homogeneous media has been investigated extensively for more than half a century \citep{clift2005bubbles},  it might  be worthwhile to extend this research towards (strongly) stratified liquids.    

\section*{Acknowledgements}
We acknowledge support from the Netherlands Center for Multiscale Catalytic Energy Conversion (MCEC), a NWO Gravitation programme funded by the Ministry of Education, Culture and Science of the government of the Netherlands, an ERC-Advanced Grant under project number 740479, the Balzan Foundation.
Y. L. acknowledges the financial support from the Fundamental Research Funds for the Central Universities and the Natural Science Foundation of China under grant No. 12272376.
C. D. kindly acknowledges financial support by the Industrial Partnership Programme (IPP) of the Netherlands Organization for Scientific Research (NWO). This research programme is co-financed by Canon Production Printing Holding B.V., University of Twente and Eindhoven University of Technology.

\section*{Declaration of Interest}
The authors report no conflict of interest. 

\appendix
\section{Physical properties of the ethanol-water mixture}\label{appA}
\begin{figure}
  \centerline{\includegraphics[width=0.85\textwidth]{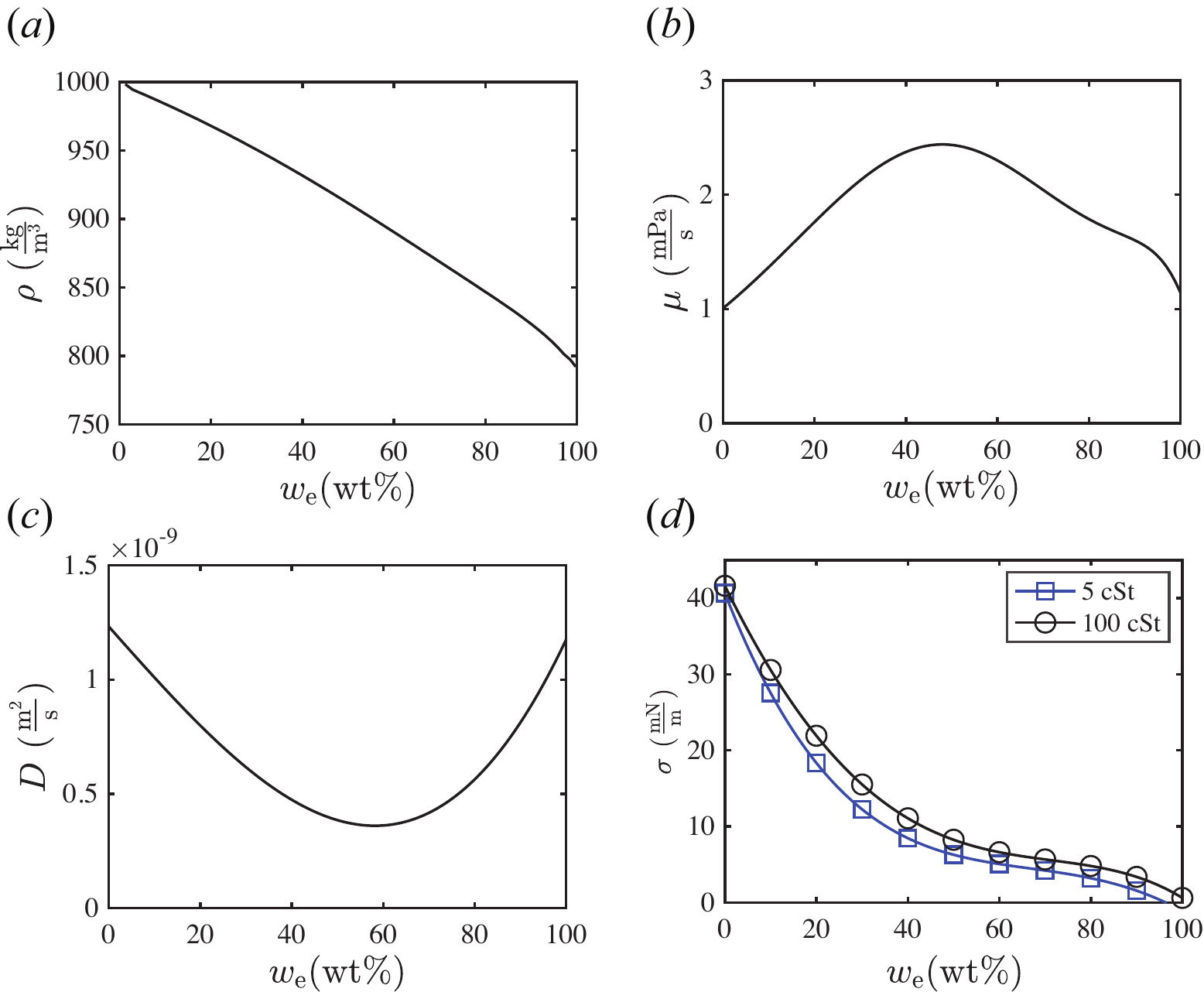}}
  \caption{Physical properties of the ethanol-water mixture for different ethanol weight fraction gradients. ($a$) Density $\rho$, ($b$) dynamic viscosity $\mu$ \citep{khattab2012density} and ($c$) diffusivity $D$ \citep{par2013mutual}, respectively.  ($d$) Interfacial surface tension $\sigma(w_{\mathrm{e}})$ between the two different silicone oils and the ethanol-water mixture.   }
\label{fig:10}
\end{figure}

Physical properties of the ethanol-water mixtures for different ethanol weight fractions are taken from \citet{khattab2012density} and \citet{par2013mutual},  see figures \ref{fig:10}($a$)-($c$).  The interfacial surface tension $\sigma(w_{\mathrm{e}})$ between the silicone oil and the ethanol-water mixture is measured using the pendant drop method on a goniometer and shown in figure \ref{fig:10}($d$).  The markers indicate the average value of six measurements, where the standard deviation is on the order of the size of the markers.

\section{Contours of the velocity magnitude}\label{appC}
\begin{figure}
  \centerline{\includegraphics[width=0.40\textwidth]{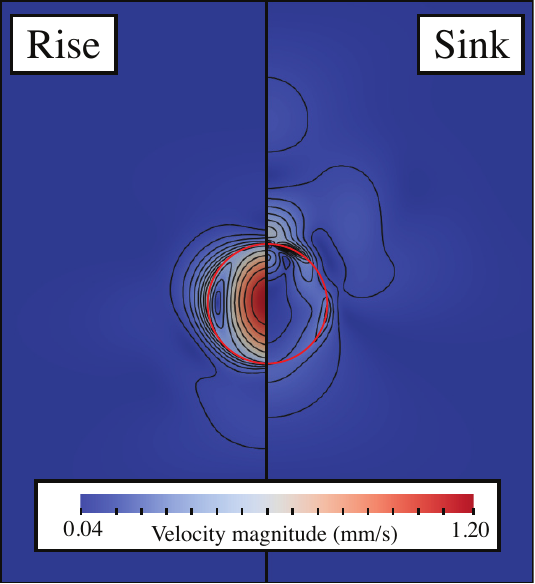}}
  \caption{Contours of the velocity magnitude in close proximity to the drop during rising (left) and sinking (right).}
\label{fig:12}
\end{figure}

Figure \ref{fig:12} shows the contours of the velocity magnitude in- and outside the drop as the drop reaches its peak velocity during rising (left) and sinking (right). The Marangoni flow is strong as the drop rises, causing the contours to be closely packed at the interface of the drop. Additionally, a strong internal circulation inside the drop is visible. On the other hand, during sinking, the Marangoni flow is very weak and the contour lines are not as closely packed. It can be seen that the internal circulation has vanished.

\bibliographystyle{jfm}
\bibliography{References.bib}

\end{document}